\documentclass[onecolumn,prd,nofootinbib,showpacs]{revtex4}

\usepackage{textcomp}
\usepackage{mathtools, ulem, graphicx}
\usepackage[export]{adjustbox}
\usepackage{sidecap}
\usepackage{verbatim}
\usepackage{natbib}

\newcommand{\e}{\begin{equation*}\begin{aligned}}
\newcommand{\ee}{\end{aligned}\end{equation*}}

\newcommand{\en}{\begin{equation}\begin{aligned}}
\newcommand{\een}{\end{aligned} \end{equation}}

\newcommand{\pfa}[2]{\frac{\delta #1}{\delta #2}}

\newcommand{\p}{\partial}
\newcommand{\pf}[2]{\frac{\p #1}{\p #2}}
\newcommand{\f}[2]{\frac{#1}{#2}}

\newcommand{\ra}{\rangle}
\newcommand{\la}{\langle}

\newcommand{\da}{\dagger}
\newcommand{\ma}{\mathcal}

\newcommand{\tr}{\text{Tr}}

\newcommand{\Q}{\left}
\newcommand{\W}{\right}

\newcommand{\pma}{\begin{pmatrix}}
\newcommand{\epma}{\end{pmatrix}}

\newcommand{\bma}{\begin{bmatrix}}
\newcommand{\ebma}{\end{bmatrix}}

\newcommand{\na}{\nabla}
\newcommand{\de}{\delta}
\newcommand{\ep}{\epsilon}
\newcommand{\om}{\omega}

\begin{document}

\title{Path-Integral Derivation of the Non-relativistic Scale Anomaly}
\author{Chris L. Lin}
\author{Carlos R. Ord\'{o}\~{n}ez}
\affiliation{Department of Physics, University of Houston, Houston, TX 77204-5005}

\date{\today}
\email{cllin@uh.edu}
\email{cordonez@central.uh.edu}


\begin{abstract}
In this paper we calculate the scale anomaly for a quantum field theoretic 2D non-relativistic Bose gas with contact interactions using Fujikawa's method, both in vacuum and in many-body systems. The use of path integrals for these problems is novel and motivated by a recently developed path-integral framework for addressing questions about scaling in these systems. A natural class of regulators is found that produces the correct value of the anomaly traditionally calculated via other methods, e.g., diagrammatically via the $\beta$ function. 
\end{abstract}

\maketitle

\section{Introduction}

The use of Fujikawa's method in particle physics is well known and is now standard in textbooks \cite{weinberg}. It was originally developed to understand the chiral anomaly \cite{fujikawa} but has since been extended to other cases, including the relativistic scale anomaly \cite{Shizuya}. However, as far as we are aware, it has not been used before for non-relativistic physics. There are currently reasons to embark in such calculations. Non-relativistic anomalies have been studied since the seminal paper by R. Jackiw \cite{jack}, mostly using canonical methods, not Fujikawa's\footnote{See \cite{6,7} and references therein.}. Interest in these anomalies has intensified in the study of ultracold 2D gases  \cite{14,15,16,17,18,19,20,22,23}, with the work by J. Hofmann on anomalies of trapped 2D Fermion gases being of particular relevance \cite{hoff}. Despite all this activity, there are still questions about anomalies and their impact in such systems that need to be answered \cite{21}. A path-integral Fujikawa approach to study anomalies in systems with an $SO(2,1)$ classical symmetry, mainly in the context of 2D diluted gases, has been recently proposed in \cite{ord}.  While this approach provides a nice picture of the structure of anomalies in many-body systems, the calculation of the Fujikawa Jacobian is crucial in order for this framework to also provide a practical scheme that will help us better understand the role of anomalies in lower-dimensional physics.   We present here our first results of the Jacobian calculation for 2D complex fields with contact interactions in the case of constant background fields.  \\

Within the path-integral formulation, anomalies result from the presence of Jacobians due to the non-invariance of the measure under symmetry transformations. These 
Jacobians are functional determinants and need to be regularized. For the chiral anomaly, all regulators lead to a finite result, whereas for the relativistic scale anomaly an infinite piece remains that is present even if the same regulator is used in the free theory, so this piece can be subtracted if the free theory is taken to be non-anomalous \cite{umezawa}. The non-relativistic scale anomaly is similar to the relativistic case in this respect. However, unlike the latter, space and time are treated on unequal footing in the former. Indeed, traditionally, for both the relativistic chiral and scale anomalies, one goes into Euclidean space where the Lagrangian kinetic operator is Hermitian. In this Euclidean space one can work with functions of a single variable (the 4 momentum squared) that is positive semi-definite in all directions. In contrast, for the non-relativistic case the Lagrangian operator is Hermitian in real time (``Minkowski space''). Due to the asymmetry between space and time, one is stuck with $\omega$ and $\vec{k}\,^2$ rather than a single $k^2$, making the task considerably more difficult, which may be a reason for why this problem has not been addressed before using Fujikawa's method. \\

The structure of this paper is as follows: we give a brief introduction to Fujikawa's method, after which we review the essential technical details for the system that will be considered here. We then proceed with the Jacobian calculation for zero and finite temperature. Conclusions and comments end the paper. 

\section{Fujikawa's Derivation}

The derivation of the anomaly via Fujikawa's method presented here follows closely  the path-integral derivation of the Ward identities, but now the Jacobian of the symmetry transformation is taken into account. Indeed, anomalies represent a breakdown of the Ward identities, and it is precisely the Jacobian that invalidates the identities. For simplicity we will demonstrate the derivation for a scalar field theory without sources: the generalization to other (multiple) fields is straightforward. With a change of variables given by $\phi'(x)=\phi(x)+\eta \delta \phi(x)$:

\en
\int [d\phi] e^{iS[\phi]}&=\int [d\phi'] \Q |\pfa{\phi}{\phi'}\W | e^{iS[\phi(\phi')]} \\
&=\int [d\phi'] \Q |\delta^d(x-y) - \eta \pfa{\delta \phi'(x)}{\phi'(y)} \W |e^{iS[\phi'-\eta \delta \phi']}\\
&=\int [d\phi] \Q |\delta^d(x-y) - \eta \pfa{\delta \phi(x)}{\phi(y)} \W |e^{iS[\phi-\eta \delta \phi]}\\
&=\int [d\phi] e^{-\eta \int d^d x  \pfa{\delta \phi}{\phi} }e^{iS[\phi]}e^{-i\eta  \int d^dx \pfa{S}{\phi}\delta \phi}\\
&=\int [d\phi] e^{iS[\phi]} \Q( 1-\eta \int d^d x  \pfa{\delta \phi}{\phi} -i \eta  \int d^dx \pfa{S}{\phi}\delta \phi  \W).
\een

Since this holds for any volume $V$, it follows:

\en
\Q\la  \pfa{S}{\phi}\delta \phi \W\ra=i\Q\la   \Q.\pfa{\delta \phi(x)}{\phi(y)}\W|_{y=x} \W\ra.
\een

Now $\pfa{S}{\phi}\delta \phi =\pf{\ma L}{\phi}\delta \phi-\p_\mu\pf{\ma L}{\p_\mu \phi}\de \phi$. However, if $\de \phi$ is a symmetry transformation, then $\pf{\ma L}{\phi}\delta \phi+\pf{\ma L}{\p_\mu \phi}\de \p_\mu\phi=\p_\mu K^\mu $, so $\pfa{S}{\phi}\delta \phi =-\pf{\ma L}{\p_\mu \phi}\de \p_\mu\phi+\p_\mu K^\mu-\p_\mu\pf{\ma L}{\p_\mu \phi}\de \phi$ or $\pfa{S}{\phi}\delta \phi=\p_\mu \Q(-\pf{\ma L}{\p_\mu \phi}\delta \phi+K^\mu \W)=-\p_\mu j^\mu$. \\

So Fujikawa's method tells us that:

\en
\Q\la  \p_\mu j^\mu \W\ra=-i\Q\la   \Q.\pfa{\delta \phi(x)}{\phi(y)}\W|_{y=x} \W\ra.
\een

Had we added a source term $\int d^dx \, J(x)\phi(x)$, the equation would read:

\en
\Q\la  \p_\mu j^\mu \W\ra -\Q   \la  J \delta \phi \W\ra  =-i\Q\la   \Q.\pfa{\delta \phi(x)}{\phi(y)}\W|_{y=x} \W\ra.
\een

Differentiation w.r.t. to $J(x_i)$ $n$ times and setting $J=0$ would create contact terms:

\en
\Q\la  \p_\mu j^\mu(x) \phi(x_1)...\phi(x_n) \W\ra +i\sum\limits_{i=1}^{n}\Q \la \phi(x_1)...\delta \phi(x_i)\de^d(x-x_i)...\phi(x_n) \W\ra  =-i\Q\la   \Q.\pfa{\delta \phi(x)}{\phi(y)}\W|_{y=x} \phi(x_1)...\phi(x_n) \W\ra.
\een

Eq. (5) without the Jacobian contribution is the traditional Ward identity at zero temperature, in vacuum, presented in most textbooks \cite{sred}. In our case, we only need the Jacobian of the infinitesimal transformation by itself in order to compute the RHS of Eq. (3) and compare our results with the literature for both the zero-temperature and the finite-temperature case. For the latter, we will work within the framework of reference \cite{ord}, for which a detailed calculation is mandatory.

\section{Contact Interaction}

The Schr\"{o}dinger Lagrangian density for bosons with contact interaction in 2D is given by:

\en
\ma L=\psi^\da \Q( i\p_t+\f{\nabla^2}{2}\W)\psi-\f{g}{2}(\psi^\da \psi)^2,
\een

which is the 2-body interaction with a $V(\vec{x}-\vec{y})=g\,\delta^2(\vec{x}-\vec{y})$ potential:

\en
\ma L=\psi^\da \Q( i\p_t+\f{\nabla^2}{2}\W)\psi-\f{1}{2} \int d^2 \vec{y} \,\psi^\da(t,\vec{x}) \psi(t,\vec{x})V(\vec{x}-\vec{y}\,)\psi^\da(t,\vec{y}) \psi(t,\vec{y}).
\een

The action corresponding to this Lagrangian is scale-invariant. This can be readily seen by noting that in $D=2$, the coupling $g$ has no dimensions in units of length (with $\hbar=m=1$). Therefore Eq. (3) applies.

\section{Scale Transformation}

Under a non-relativistic dilation transformation \cite{jackiw}:

\begin{flalign}
&\vec{x}\,'=\lambda \vec{x}, \\ \nonumber
&t'=\lambda^2 t,\\ \nonumber
&\psi'(\vec{x}\,' ,t')=\lambda^{-D/2} \psi(\vec{x},t). \nonumber
\end{flalign}

Setting $\lambda=1+\eta$ for infinitesimal $\eta$:

\en
\de \vec{x} &=\eta \vec{x}, \\
\de t &=2\eta t,\\
\tilde{\de} \psi &=\eta \theta  \psi(t,\vec{x})\equiv\eta \delta\psi,\\
\tilde{\de} \psi^* &=\eta \theta  \psi^*(t,\vec{x})\equiv \eta \delta \psi^*,\\
\theta &\equiv \Q(-\f{D}{2}-\vec{x}\cdot \vec{\na}-2t \p_t \W).
\een

where $D=d-1$ is the spatial dimension\footnote{We used $\tilde{\de} \psi$ ($\tilde{\de} \psi^*$) for the infinitesimal change in $\psi$ ($\psi^*$), and set $\tilde{\de} \psi=\eta \de\psi$ to make the notation consistent with Eq. (2).}. In this paper we will set $D=2$. Therefore\footnote{Sometimes we write $x=(x_0,\vec{x})=(t,\vec{x})$ for notational convenience.}:

\en
 \Q.\pfa{\delta \psi(x)}{\psi(y)}\W|_{y=x}=\Q.[ \theta \delta(x_0-y_0)\delta^2(\vec{x}-\vec{y})]\W|_{y=x}=\Q.\pfa{\delta \psi^*(x)}{\psi^*(y)}\W|_{y=x}.
\een


Note that unlike translations, for dilations both conventions --- $\de \psi=\eta \Q(-1-\vec{x}\cdot \vec{\na}-2t \p_t \W) \psi(t,\vec{x})$ or $\de \psi=\eta \Q(1+\vec{x}\cdot \vec{\na}+2t \p_t \W) \psi(t,\vec{x})$ --- leading to currents of opposite sign, are widely used. We've adopted the former, which leads to a dilation charge of \cite{hagen}:

\en
D=\int d^2 \vec{x} \, \vec{x}\cdot \vec{j}-2tH,\\
\vec{j}=-\f{i}{2} \Q( \psi^\da \vec{\nabla} \psi-\vec{\nabla}\psi^\da  \psi \W).
\een


\section{Fujikawa Calculation: Set Up}

The generalization of the scalar case to our Lagrangian is straightforward:

\en
\det 
\pma
\pfa{\psi(x)}{\psi'(y)} & \pfa{\psi(x)}{\psi'^{*}(y)}\\
\pfa{\psi^*(x)}{\psi'(y)} & \pfa{\psi^*(x)}{\psi'^{*}(y)}
\epma&=
\det
\pma
\delta^3(x-y)-\eta \theta \delta^3(x-y) & 0\\
0 & \delta^3(x-y)-\eta \theta \delta^3(x-y)
\epma \\
&=\exp \Q(- \eta \int dtd^2\vec{x} \, \text{tr} 
\Q.\bma
\theta \delta(x_0-y_0)\delta^2(\vec{x}-\vec{y}) & 0\\
0 & \theta \delta(x_0-y_0)\delta^2(\vec{x}-\vec{y})
\ebma \W|_{y=x}
\W).
\een

where we've used $\det A=e^{\hat{\text{T}}\text{r}\log A}$ \footnote{$\hat{\text{T}}\text{r}$ includes \textit{both} functional and matrix indices; $\text{tr}$ only refers to the $2\times 2$ matrix indices in Eqs. (12) and (13).}. Comparison with Eq. (3) makes the generalization clear:

\en
\Q\la  \p_\mu j^\mu \W\ra=-i \text{tr} 
\Q.\bma
\theta \delta^3(x-y) & 0\\
0 & \theta\delta^3(x-y)
\ebma \W|_{y=x}.
\een

This expression is singular so needs to be regularized. This is done by expanding $\delta^3(x-y) I_2$, where $I_2=\pma 1 & 0 \\0 & 1 \epma$, using the eigenbasis $\phi_n$ of a Hermitian operator $M$:

\en
\delta^3(x-y) I_2=\sum \limits_{n}\phi_n(x_0,\vec{x})\phi_n^\da(y_0,\vec{y}).
\een

Inserting a regulator that's a function of $M$ 

\en
\delta^3_R(x-y) I_2=\sum \limits_{n} R\Q(\f{M}{\Lambda^2}\W)\phi_n(x_0,\vec{x})\phi_n^\da(y_0,\vec{y}),
\een

with the property that $R(0)=1$ so that at the end of the calculation we send $\Lambda \rightarrow \infty$ and $\lim \limits_{\Lambda \rightarrow \infty}R\Q(\f{M}{\Lambda^2}\W)=1$ ($[M]=[\Lambda^2]$). The idea is to choose $R$ such that large eigenvalues are suppressed giving a convergent sum:

\en
\delta^3_R(x-x) I_2=\sum \limits_{n} R\Q(\f{\lambda_n}{\Lambda^2}\W)\phi_n(x_0,\vec{x})\phi_n^\da(x_0,\vec{x}).
\een

Once the Hermitian operator $M$ has been selected, then the sum over $n$ in Eq. (15) gives:

\en
\delta^3_R(x-y) I_2&=\sum \limits_{n}R\Q(\f{M}{\Lambda^2}\W)\phi_n(x_0,\vec{x})\phi_n^\da(y_0,\vec{y})\\
&= R\Q(\f{M}{\Lambda^2}\W)\delta^3(x-y) I_2,
\een

so that 

\en
\text{tr} \Q[\delta^3_R(x-y) I_2\W]=\text{tr} \Q[ R\Q(\f{M}{\Lambda^2}\W)\delta^3(x-y) I_2\W].
\een

For the class of regulators defined by Eqs. (19) and (20) in the next section - a consistent choice for the untrapped system - it will be shown that the non-trivial contribution to Eq. (18) will have an even integrand in \textit{both} $\omega$ and $\vec{k}$ (Eq. (28)). This means that the derivative terms in $\theta$ will give a null contribution and the only term that will survive is given by Eq. (18). To see this, one should take the space-time derivatives in Eq. (21), and then set $x=y$; these terms will give odd contributions to the integrand in $(\omega, \vec{k})$ space when multiplied by the even terms of Eq. (28). \\

Therefore, Eq. (18) is the expression we aim to calculate. 

\section{Fujikawa Calculation: Mode Expansion}

\subsection{Zero Temperature}

For our Hermitian matrix we will choose \cite{bbs}:

\en
M=\begin{pmatrix}
i\p_t+\f{\nabla^2}{2}+\mu-2g\psi^* \psi+i\ep & -g \psi^2 \\
-g \psi^{*2} & -i\p_t+\f{\nabla^2}{2}+\mu-2g\psi^* \psi+i\ep
\end{pmatrix}
\een

where the fields in $M$ are constant background fields, and $\f{1}{2}\pma \chi^\da & \chi \epma M \pma \chi \\ \chi^\da \epma$ is the quadratic Lagrangian resulting from a saddle point expansion of the action about the background field $\psi$, and $\chi$ is the shift of the original field from $\psi$. We've included a chemical potential $\mu$ for many-body physics that explicitly breaks scale-invariance, but we can always set $\mu=0$ and as we will demonstrate, the inclusion of $\mu$ has no effect on the anomaly. \\

For our regulating function $R$ we will choose 

\en
R\Q(\f{M}{\Lambda^2}\W)=\Q( 1 \pm \f{M}{\Lambda^2}\W)^{-1},
\een

which clearly satisifies $R(0)=1$. Plugging in this regulator into Eq. (17) and Fourier expanding $\delta^3(x-y)$ gives:

\en
\text{tr} \Q[\delta^3_R(x-y) I_2\W]=\int \f{d\omega}{2\pi} \int \f{d^2\vec{k}}{(2\pi)^2} 
\text{tr}\begin{pmatrix}
1\pm \f{\omega-\f{k^2}{2}+\mu-2g \psi^*\psi+i\epsilon} {\Lambda^2} & \mp \f{g\psi^2}{\Lambda^2} \\
\mp \f{g\psi^{*2}}{\Lambda^2}&1\pm\f{-\omega-\f{k^2}{2}+\mu-2g \psi^*\psi+i\epsilon}{\Lambda^2}
\end{pmatrix}^{-1}e^{-i\om (x_o-y_o)+i\vec{k}\cdot (\vec{x}-\vec{y})}.
\een

We will now take $(x_0,\vec{x})=(y_0,\vec{y})$; will make a change of variables $\tilde{\omega}=\f{\omega}{\Lambda^2}$ and $\tilde{k}=\f{k}{\Lambda}$; and then replace the tildes since they are dummy indices ($\omega$ and $\vec{k}$ are now dimensionless):

\en
\text{tr}\Q[\delta_R(0)I_2\W]=\Lambda^4\int \f{d\omega}{2\pi} \int \f{d^2k}{(2\pi)^2}
\text{tr}\begin{pmatrix}
1\pm \Q(\omega-\f{k^2}{2}+\f{\mu-2g \psi^*\psi+i\epsilon } {\Lambda^2}\W) & \mp \f{g\psi^2}{\Lambda^2} \\
 \mp \f{g\psi^{*2}}{\Lambda^2}&1\pm\Q(-\omega-\f{k^2}{2}+\f{\mu-2g \psi^*\psi+i\epsilon}{\Lambda^2}\W) \\
\end{pmatrix}^{-1}.
\een

For notational convenience we will write the above expression as:

\en
\text{tr}\Q[\delta_R(0)I_2\W]=\pm\Lambda^4\int \f{d\omega}{2\pi} \int \f{d^2 \vec{k}}{(2\pi)^2} \,
\text{tr}\begin{pmatrix}
\omega-\f{k^2}{2}+A_\pm+i\ep & -\f{g\psi^2}{\Lambda^2} \\
 - \f{g\psi^{* 2}}{\Lambda^2}&-\omega-\f{k^2}{2}+A_\pm+i\ep \\
\end{pmatrix}^{-1}, 
\een

with 

\en
A_\pm=\pm 1 +\f{\mu-2g \psi^*\psi}{\Lambda^2}.
\een

To evaluate the inverse in Eq. (23), we will use the identity for matrix inverses $(D+B)^{-1}=D^{-1}-(D^{-1}B)D^{-1}+(D^{-1}B)(D^{-1}B)D^{-1}-...$ with 

\en
D_\pm&=\pma \omega-\f{k^2}{2}+A_\pm+i\ep &  0 \\ 0 & -\omega-\f{k^2}{2}+A_\pm +i\ep \epma , \\
B&=\pma 0 &  -\f{g\psi^2}{\Lambda^2} \\ -\f{g\psi^{*2}}{\Lambda^2} & 0 \epma .
\een

Note that the $i\epsilon$ makes $D_\pm$ invertible. \\

So Eq. (22) becomes:

\en
\text{tr}\Q[\delta_R(0)I_2\W]=\pm\Lambda^4\int \f{d\omega}{2\pi} \int \f{d^2\vec{k}}{(2\pi)^2}\, 
\text{tr}\Q(D^{-1}_\pm-(D^{-1}_\pm B)D^{-1}_\pm+(D^{-1}_\pm B)(D^{-1}_\pm B)D^{-1}_\pm    \W),
\een

where we terminated the series at two powers of $B$, since each additional power of $B$ produces a $\f{1}{\Lambda^2}$ that the $\Lambda^4$ prefactor can't offset.\\

The first term in the series, $D^{-1}_\pm$, when doing the integral over $\omega$, is independent of the coupling:

\en
\pm \Lambda^4\int \f{d\omega}{2\pi} \int \f{d^2\vec{k}}{(2\pi)^2}
\text{tr}\begin{pmatrix}
\f{1}{\omega-\f{k^2}{2}+A_\pm+i\epsilon} & 0 \\
0&\f{1}{-\omega-\f{k^2}{2}+A_\pm+i\epsilon} \\
\end{pmatrix} \\
=\pm\Lambda^4\int \f{d\omega}{2\pi} \int \f{d^2 \vec{k}}{(2\pi)^2} \Q(\f{2 (\f{k^2}{2}-A_\pm-i\epsilon)}{\omega^2-(\f{k^2}{2}-A_\pm-i\epsilon)^2} \W)\\
=\f{\mp i}{2}\Lambda^4 \int \f{d^2 \vec{k}}{(2\pi)^2}. 
\een 

 Therefore this term is also contained in the free-case, which we take to be anomaly-free. So we subtract this term when calculating the anomaly. The next term $(D^{-1}_\pm B)D^{-1}_\pm$ has no diagonal elements, so is traceless. \\

The only term to calculate is the $(D^{-1}_\pm B)(D^{-1}_\pm B)D^{-1}_\pm$ term which produces:

\en
\text{tr}\Q[\delta_R(0)I_2\W]&=\pm \Lambda^4\int \f{d\omega}{2\pi} \int \f{d^2\vec{k}}{(2\pi)^2}  \Q(-\f{g\psi^2}{\Lambda^2}\W) \Q(-\f{g\psi^{* 2}}{\Lambda^2}\W)
\Q(\f{-2 (\f{k^2}{2}-A_\pm-i\epsilon)}{\Q[\omega^2-(\f{k^2}{2}-A_\pm-i\epsilon)^2\W]^2} \W)\\
&=\pm g^2 (\psi^* \psi)^2 \int \f{d\omega}{2\pi} \int_0^\infty \f{dk}{2\pi} \Q(\f{-2k (\f{k^2}{2}-A_\pm-i\epsilon)}{\Q[\omega^2-(\f{k^2}{2}-A_\pm-i\epsilon)^2\W]^2} \W).
\een

The integral over $k$ is straightfoward:

\en
\text{tr}\Q[\delta_R(0)I_2\W]=\pm \f{g^2 (\psi^* \psi)^2}{2\pi} \int \f{d\omega}{2\pi}  \Q(\f{1}{\omega^2-(A_\pm+i\epsilon)^2} \W).
\een 

Now $A_\pm$ in Eq. (24) can be safely taken to $\pm 1$ ($\Lambda \rightarrow \infty$). For both $\pm$ cases, the result is the same:

\en
\text{tr}\Q[\delta_R(0)I_2\W]=i\f{g^2 (\psi^* \psi)^2}{4\pi}. 
\een 

Plugging this into Eq. (13) gives:

\en
\Q\la  \p_\mu j^\mu \W\ra=-\f{g^2 (\psi^* \psi)^2}{4\pi}. 
\een

This can be compared with \cite{bergman} (for the case of constant background fields) by making the replacement $g \rightarrow \f{g}{2}$. \\

Because both $R\Q(\f{M}{\Lambda^2}\W)=\Q( 1 \pm \f{M}{\Lambda^2}\W)^{-1}$ work as regulators, any linear combination such that their coefficients add to one works. For example:

\en
R\Q(\f{M}{\Lambda^2}\W)&=\f{1}{2}\Q( 1 + \f{M}{\Lambda^2}\W)^{-1}+\f{1}{2}\Q( 1 - \f{M}{\Lambda^2}\W)^{-1}\\
&=\Q(1-\f{M^2}{\Lambda^4}\W)^{-1}.
\een

We have also verified that the following regulators work: 

\en
R\Q(\f{M}{\Lambda^2}\W)=\Q( 1 \pm \f{M}{\Lambda^2}\W)^{-2}.
\een

\subsection{Many-Body} 

Under the formalism developed in \cite{ord}\footnote{The factor of $\beta A$ in the denominator of Eq. (52) in \cite{ord} cancels the Euclidean version of the factor $\int_0^\beta d\tau \int d^2 \vec{x}$ from Eq. (12) in this paper, since for constant background fields our class of regulators gives a constant value for $\text{tr}\Q(\delta_R(0)I_2\W)$. The time and spatial derivatives in $\hat{\theta}_s$ and $\hat{\theta}$ in paper \cite{ord} give no contributions in this case (untrapped) as explained here. Notice Eq. (34) is a $2\times 2$ version of Eq. (52) in  \cite{ord}.}:

\en
2\ma E-2P=\text{tr}\Q[\delta_R(0)I_2\W].
\een

However, here the anomalous term is evaluated in Euclidean space using the finite temperature rules. Eq. (28) with plus chosen is:

\en
\text{tr}\Q[\delta_R(0)I_2\W]=\f{g^2 (\psi^* \psi)^2}{2\pi} \int \f{d\omega}{2\pi}  \Q(\f{1}{\omega^2-(1+i\epsilon)^2} \W).
\een 

In terms of the original dimensionful $\omega$:

\en
\text{tr}\Q[\delta_R(0)I_2\W]&=\f{g^2 (\psi^\da \psi)^2}{2\pi} \Lambda^2 \int \f{\Lambda^2 d\omega}{2\pi}  \Q(\f{1}{(\Lambda^2 \omega)^2-(\Lambda^2+i\epsilon)^2} \W) \\
&=\f{g^2 (\psi^\da \psi)^2}{2\pi} \Lambda^2 \int \f{d\omega}{2\pi}  \Q(\f{1}{\omega^2-(\Lambda^2+i\epsilon)^2} \W).
\een 

When going to finite temperature, the difference is that we have $-\p_\tau$ instead of $i\p_t$. The effect is to replace $\omega$ with $i\omega$ in Eq. (35). That is, had we started directly in Euclidean space, we would still get Eq. (35), but with $\omega$ replaced by $i\omega$ stemming from $-\p_\tau$ replacing $i\p_t$ in our regulator. The second change is that the integral is a sum since the modes are discrete, with a $\beta$ factor resulting from writing the delta function as $\delta^d(x-y)=\f{1}{\beta}\sum\limits_{n}\int \f{d^Dk}{(2\pi)^D}e^{-i\omega_n(x_0-y_0)}e^{i\vec{k}\cdot(\vec{x}-\vec{y})}$. So a sum replaces the integral. So had we started directly in Euclidean space, we would have a sum over frequencies instead of an integral:

\en
\text{tr}\Q[\delta_R(0)I_2\W]=\f{g^2 (\psi^* \psi)^2}{2\pi} \f{\Lambda^2}{\beta}\sum\limits_{n}  \Q(\f{1}{-\omega_n^2-(\Lambda^2+i\epsilon)^2} \W),
\een 

where $\omega_n=\f{2\pi n}{\beta}$ for bosons. The $i\ep$ no longer matters and the summation is standard:

\en
\text{tr}\Q[\delta_R(0)I_2\W]=-\f{g^2 (\psi^* \psi)^2}{2\pi} \f{\Lambda^2}{\beta} \left( \f{\beta \coth\Q(\f{\beta \Lambda^2}{2}\W)}{2\Lambda^2}\right),
\een 

which in the limit of large $\Lambda$ gives:

\en
\text{tr}\Q[\delta_R(0)I_2\W]=-\f{g^2 (\psi^* \psi)^2}{4\pi}. 
\een 

So plugging this into Eq. (34)

\en
2\ma E-2P=-\f{g^2 (\psi^* \psi)^2}{4\pi}, 
\een

which agrees with \cite{norway} with $g \rightarrow 2g$. For the finite temperature case, there is some ambiguity in the continuation to Euclidean space that affects the sign, where $A_+$ leads to the correct sign, and $A_-$ leads to the negative sign. We take the view that the zero-temperature limit must reproduce the vacuum result.

\section{Conclusion}

Fujikawa's path-integral method has been applied to the Schr\"{o}dinger Lagrangian to describe anomalies for 2D non-relativistic, $SO(2,1)$ scale-invariant complex bosons with contact interactions. A class of natural regulators was identified that gives results consistent with those in the literature, obtained with other methods, in both zero and finite-temperature cases \cite{bergman, norway}. This work was motivated by the recent formulation of Fujikawa's approach to analyze the anomaly structure for 2D gases with $SO(2,1)$ classical symmetry (and other systems with such symmetry) \cite{ord}, which is relevant in the study of ultracold 2D trapped gases \cite{hoff}. It was important, therefore, that we made contact with established work using other techniques. Further work is needed for a deeper understanding of this method and its possible applications. In particular, heat kernel techniques will be used to investigate trapped systems. Work on these issues is in progress \cite{prog}.


\end{document}